\documentclass[nohyper,12pt,letterpaper]{JHEP3}
\usepackage{epsfig}
\usepackage[latin1]{inputenc}
\usepackage{bbm,amsfonts}
\usepackage{graphicx}
\usepackage{amssymb,amsmath}

\author{Marco S. Bianchi$^\ast$,
  Matias Leoni$^{\ast}$,
  Andrea Mauri$^{\dag,\hash}$,
  Silvia Penati$^\ast$
  and Alberto Santambrogio$^{\hash}$\\\\
  $^\ast$Dipartimento di Fisica, Universit\`a di Milano--Bicocca and
  INFN, Sezione di Milano--Bicocca, Piazza della Scienza 3, I-20126 Milano, Italy \\\\
  $^\dag$Dipartimento di Fisica dell'Universit\`a degli studi di Milano\\\\
  $^\hash$ INFN, Sezione di Milano, via Celoria 16, I-20133 Milano, Italy
  \qquad\\\\
  E-mail: \email{marco.bianchi@mib.infn.it, matias.leoni@mib.infn.it,
    andrea.mauri@mi.infn.it, silvia.penati@mib.infn.it,
    alberto.santambrogio@mi.infn.it }}

\abstract{For ${\cal N}=6$ superconformal Chern--Simons--matter theories  in three dimensions, by a direct superspace Feynman diagram approach,
we compute the two--loop four--point scattering amplitude with external chiral matter fields. We find that the result is in perfect agreement with the two--loop result for a light--like four--polygon Wilson loop. This is a 
nontrivial evidence of the scattering amplitudes/Wilson loop duality in three dimensions.
Moreover, both the IR divergent and the finite parts of our two--loop result agree with a BDS--like ansatz for all--loop amplitudes where the scaling function is given in terms of the ${\cal N}=4$ SYM one, according to
the conjectured Bethe equations for ABJM. Consequently, we are able to make a prediction for the four--loop correction to the amplitude. We also discuss the dual conformal invariance of the two--loop
result.}

\preprint{July 2011\\ IFUM-480-FT}

\title{SCATTERING AMPLITUDES/WILSON LOOP DUALITY IN ABJM THEORY}


\csname @addtoreset\endcsname{equation}{section}

\def\bseq{\begin{subequation}}  
\def\eseq{\end{subequation}}
\def\bsea{\begin{subeqnarray}}  
\def\esea{\end{subeqnarray}}

\newcommand{\beq}{\begin{equation}}
\newcommand{\bea}{\begin{eqnarray}}
\newcommand{\eea}{\end{eqnarray}}
\newcommand{\eeq}{\end{equation}}

\newcommand {\non}{\nonumber}

\renewcommand{\a}{\alpha}
\renewcommand{\b}{\beta}

\renewcommand{\d}{\delta}

\newcommand{\G}{\Gamma}

\newcommand{\e}{\epsilon}

\renewcommand{\l}{\lambda}

\newcommand{\s}{\sigma}

\newcommand{\Db}{\overline{D}}

\def\Mb{\kern 2pt\mathchoice
        {
         \vbox{\hrule width10pt height 0.4pt depth 0pt
         \kern 1.2pt\hbox{\kern -2pt$\displaystyle M$}}}
        {
         \vbox{\hrule width10pt height 0.4pt depth 0pt
         \kern 1.2pt\hbox{\kern -2pt$\textstyle M$}}}
        {
\vbox{\hrule width6pt height 0.4pt depth 0pt
         \kern 1.0pt\hbox{\kern -2pt$\scriptstyle M$}}}
        {
         \vbox{\hrule width5pt height 0.4pt depth 0pt
         \kern 0.8pt\hbox{\kern -2pt$\scriptscriptstyle M$}}}}

\def\Sb{\kern 2pt\mathchoice
        {
         \vbox{\hrule width6pt height 0.4pt depth 0pt
         \kern 1.2pt\hbox{\kern -2pt$\displaystyle S$}}}
        {
         \vbox{\hrule width6pt height 0.4pt depth 0pt
         \kern 1.2pt\hbox{\kern -2pt$\textstyle S$}}}
        {
         \vbox{\hrule width3.5pt height 0.4pt depth 0pt
         \kern 1.0pt\hbox{\kern -2pt$\scriptstyle S$}}}
        {
         \vbox{\hrule width3pt height 0.4pt depth 0pt
         \kern 0.8pt\hbox{\kern -2pt$\scriptscriptstyle S$}}}}

\def\Rb{\kern 2pt\mathchoice
        {
         \vbox{\hrule width5.5pt height 0.4pt depth 0pt
         \kern 1.2pt\hbox{\kern -2.5pt$\displaystyle R$}}}
        {
         \vbox{\hrule width5.5pt height 0.4pt depth 0pt
         \kern 1.2pt\hbox{\kern -2.5pt$\textstyle R$}}}
        {
         \vbox{\hrule width3.5pt height 0.4pt depth 0pt
         \kern 1.0pt\hbox{\kern -2.2pt$\scriptstyle R$}}}
        {
         \vbox{\hrule width3pt height 0.4pt depth 0pt
         \kern 0.8pt\hbox{\kern -2.2pt$\scriptscriptstyle R$}}}}

  \def\pp{{\mathchoice
      {
          \kern 1pt%
          \raise 1pt
          \vbox{\hrule width5pt height0.4pt depth0pt
            \kern -2pt
            \hbox{\kern 2.3pt
              \vrule width0.4pt height6pt depth0pt
              }
            \kern -2pt
            \hrule width5pt height0.4pt depth0pt}%
            \kern 1pt
       }
        {
          \kern 1pt%
          \raise 1pt
          \vbox{\hrule width4.3pt height0.4pt depth0pt
            \kern -1.8pt
            \hbox{\kern 1.95pt
              \vrule width0.4pt height5.4pt depth0pt
              }
            \kern -1.8pt
            \hrule width4.3pt height0.4pt depth0pt}%
            \kern 1pt
        }
        {
          \kern 0.5pt%
          \raise 1pt
          \vbox{\hrule width4.0pt height0.3pt depth0pt
            \kern -1.9pt  
            \hbox{\kern 1.85pt
              \vrule width0.3pt height5.7pt depth0pt
              }
            \kern -1.9pt
            \hrule width4.0pt height0.3pt depth0pt}%
            \kern 0.5pt
        }
        {
          \kern 0.5pt%
          \raise 1pt
          \vbox{\hrule width3.6pt height0.3pt depth0pt
            \kern -1.5pt
            \hbox{\kern 1.65pt
              \vrule width0.3pt height4.5pt depth0pt
              }
            \kern -1.5pt
            \hrule width3.6pt height0.3pt depth0pt}%
            \kern 0.5pt
        }
    }}

  \def\mm{{\mathchoice
               {
                 \kern 1pt
           \raise 1pt    \vbox{\hrule width5pt height0.4pt depth0pt
                  \kern 2pt
                  \hrule width5pt height0.4pt depth0pt}
                 \kern 1pt}
               {
                \kern 1pt
           \raise 1pt \vbox{\hrule width4.3pt height0.4pt depth0pt
                  \kern 1.8pt
                  \hrule width4.3pt height0.4pt depth0pt}
                 \kern 1pt}
               {
                \kern 0.5pt
           \raise 1pt
                \vbox{\hrule width4.0pt height0.3pt depth0pt
                  \kern 1.9pt
                  \hrule width4.0pt height0.3pt depth0pt}
                \kern 1pt}
               {
               \kern 0.5pt
         \raise 1pt  \vbox{\hrule width3.6pt height0.3pt depth0pt
                  \kern 1.5pt
                  \hrule width3.6pt height0.3pt depth0pt}
               \kern 0.5pt}
               }}

\def\pd{{\kern0.5pt
           + \kern-5.05pt \raise5.8pt\hbox{$\textstyle.$}\kern
0.5pt}}

\def\pmd{{\kern0.5pt
          \pm \kern-5.05pt
\raise6.3pt\hbox{$\textstyle.$}\kern1.5pt}}

\def\md{{\mathchoice
   {
      {{\kern 1pt - \kern-6.2pt \raise5pt\hbox{$\textstyle.$}\kern
1pt}}}
    {
      {{\kern 1pt - \kern-6.2pt \raise5pt\hbox{$\textstyle.$}\kern
1pt}}}
    {
      {\kern0.5pt - \kern-5.05pt
\raise3.4pt\hbox{$\textstyle.$}\kern0.5pt}}
    {
      {\kern0.5pt - \kern-5.05pt
\raise3.4pt\hbox{$\textstyle.$}\kern0.5pt}}}}

\def\beq{\begin{equation}}
\def\eeq{\end{equation}}
\def\bea{\begin{eqnarray}}
\def\eea{\end{eqnarray}}
\def\Tr{\textstyle{Tr}}

\def\a{\alpha}
\def\b{\beta}

\def\d{\delta}
\def\e{\epsilon}

\def\th{\theta}
\def\l{\lambda}
\def\G{\Gamma}

\begin{document}

\section{Introduction}

The development of efficient techniques for computing scattering amplitudes in gauge theories has led to the discovery of new unexpected properties in the on--shell sector of these theories.

In four dimensions, the
use of stringy--inspired methods \cite{dixon}, twistor string theory \cite{witten} and AdS/CFT
correspondence \cite{maldacena}
has allowed to dig out hidden symmetries for the planar sector of the maximally supersymmetric Yang--Mills theory. In particular, planar ${\cal N}=4$ SYM  has been proved to be integrable \cite{integrability} and the related Yangian symmetry \cite{Drummond:2009fd} to be responsible for a duality between scattering amplitudes and Wilson loops (WL).
This duality has been checked perturbatively in many cases \cite{Drummond:2006rz}--\cite{Belitsky:2011zm}, while at strong coupling
it relies on the self--duality of type IIB string on ${\rm AdS}_5 \times {\rm S}_5$ under a suitable combination of bosonic and fermionic T--dualities \cite{AM, BM}.

Another duality has been found at weak coupling which involves WL and correlation functions of BPS operators \cite{AEKMS}--\cite{Adamo:2011dq}.

Since AdS/CFT has been playing a fundamental role in the discovery of these new hidden properties
and at the same time their perturbative confirmation represents a non--trivial test of the correspondence, it is mandatory to investigate whether similar properties emerge in other
classes of theories for which a string dual description is known.

We are interested in the class of three dimensional ${\cal N}=6$ ABJM theories
\cite{ABJM} which are dual to type IIA string theory on ${\rm AdS}_4 \times
{\rm CP}_3$. A distinguished feature of these models compared to the more famous
${\cal N}=4$ SYM in four dimensions, is that they are not maximally supersymmetric.
Moreover,  the proof of the amplitudes/WL duality in type IIA string
on ${\rm AdS}_4 \times {\rm CP}_3$  is complicated by the emergence of singularities
in the fermionic T--transformations \cite{ADO}--\cite{DO}.
Therefore, a priori, it is not totally obvious that we should expect dualities and hidden symmetries to
be realized in ABJM models exactly in the same way as in their four--dimensional counterpart.

Preliminary results can be found in literature, concerning integrability \cite{MZ, GV, LMMSSST,MOS} and related Yangian symmetry \cite{BLM, Lee}. Perturbative investigation of these properties have been performed.
At tree level, scattering amplitudes are invariant under dual
superconformal symmetry \cite{HL, GHKLL} whose generators are the
level--one generators of a Yangian symmetry \cite{BLM}.
A first indication of the duality between scattering amplitudes and
WL comes from the fact that at one--loop, both the  four--point
amplitude \cite{ABM} and the light--like four--polygon WL \cite{HPW, BLMPRS} vanish.
Recently, $n$--point correlators of BPS scalar operators have been proved to vanish at one--loop
 \cite{BLMPRS}, so providing first evidence of a triad correlation functions/WL/amplitudes duality in three dimensions.

However, non--trivial perturbative support to these dualities can come only at orders
where these quantities do not vanish.

At two loops, for the ABJM model the light--like four--polygon WL has been computed in the planar limit
\cite{HPW}. In dimensional regularization, taking the light--like limit $(x_i - x_{i+1})^2 \equiv x_{i,i+1}^2 \to 0$, the result is non--vanishing and given by \footnote{This result differs from the one in the published version of Ref. \cite{HPW} by an overall minus sign and a different constant $K$. The authors of Ref. \cite{HPW} agree with us on the correctness of result (\ref{WL}).} 
\beq
\langle W_4 \rangle^{(2)}=  \l^2 \left[ - \frac{(-\mu_{WL}^2 x_{13}^2)^{2\e}}{(2\e)^2}
 - \frac{(-\mu_{WL}^2 x_{24}^2)^{2\e}}{(2\e)^2} + \frac12 \ln^2 {\left( \frac{x_{13}^2}{x_{24}^2} \right)} + C + {\cal O}(\e)
\right]
\label{WL}
\eeq
where $\mu_{WL}$ is the (properly rescaled) UV mass scale of dimensional regularization, $\l \equiv N/K$ is the ABJM coupling constant and $C = \pi^2/2 + 2 \ln{2} + 5\ln^2{2} -a_6/4$, with $a_6$ a numerical constant.

In this paper, using ${\cal N}=2$ superspace description and a direct Feynman diagram approach, we evaluate the two--loop contribution to the planar scattering superamplitude of four chiral superfields which, in components, gives rise to the amplitude for
two scalars and two chiral fermions.
The amplitude involves two external particles in the bifundamental representation of the  $U(N) \times U(N)$ gauge group and two particles in the antibifundamental. This is what mostly resembles a MHV amplitude in four dimensions.
Defining ${\cal M}_4$ to be the superamplitude divided by its tree level contribution, we find
\begin{equation}\label{amplitude}
\mathcal{M}^{(2)} \equiv \frac{\mathcal{A}_4^{(2 \, loops)}}{\mathcal{A}^{tree}_4} = \lambda^2\, \left[-\frac{( s/\mu'^2)^{-2\epsilon}}{(2\, \epsilon)^2}-\frac{(t/\mu'^2)^{-2\epsilon}}{(2\, \epsilon)^2}+\frac12\,\ln^2 \left(\frac{s}{t}\right)+{\cal C}+
\mathcal{O}(\epsilon)\right]
\end{equation}
where $\mu'$ is the (conveniently redefined) IR scale of dimensional regularization and ${\cal C}=4\zeta_2+3\ln^2 2$ is a numerical constant.

This result has a number of remarkable properties. First of all, as in the ${\cal N}=4$ SYM case, the two--loop amplitude is proportional to the tree level contribution times a function of the kinematic invariants. We find that, up to an additive, scheme dependent constant, this function matches exactly the result (\ref{WL}) once
the IR regularization is formally identified with the UV one and the particle momenta are expressed in terms of dual coordinates, $p_i = x_{i,i+1}$ (note also that the invariants in (\ref{WL}) differ from those in (\ref{amplitude}) by a sign, since the former was worked out in Minkowskian signature, whereas our result has been derived using the Euclidean metric).

Therefore, at least for the four--point amplitude, we find evidence for the following identity
\beq
\ln{ {\cal M}_4} = \ln \langle W_4 \rangle + {\rm const.}
\eeq	
that should hold order by order in the perturbative expansion of the two objects.

Quite remarkably, the two--loop result we have found has the same functional structure as the one--loop
correction to the four--point scattering amplitude for the four dimensional ${\cal N}=4$ SYM theory.
As proved in \cite{Drummond:2007aua}, for ${\cal N}=4$ SYM all momentum integrals up to four loops are dual to four dimensional true conformally invariant integrals, well defined off--shell. As a consequence, the four--point amplitude satisfies anomalous Ward identities associated to dual conformal transformations \cite{Drummond:2007au}, as dual conformal invariance is broken in the on--shell limit by the appearance of IR divergences which require introducing a mass regulator.

A natural question arises whether the same pattern is present in three dimensional ABJM theories.
We briefly discuss dual conformal invariance of the momentum integrals that occur in our two--loop diagrammatic calculation, which does not assume dual conformal invariance a priori. 
At the level of the integrands every single diagram does not appear to be invariant under dual conformal invariance, since they transform non--trivially under inversion.
Nevertheless we still expect this symmetry to be present in the on--shell amplitude, since our result matches the Wilson loop computation, which possesses the standard conformal invariance of the ABJM theory (eventually broken anomalously by UV divergences).
This means that, on--shell, it should be possible to rewrite the amplitude as a linear combination of scalar integrals which are dual invariant in three dimensions.

In the ${\cal N}=4$ SYM case, an ansatz for all--loop $n$--point
MHV amplitudes has been proposed \cite{BDS, Bern:2008ap}, where the all--loop amplitudes exponentiate and turn out to be determined by the one--loop result times the perturbative expansion of the scaling function $f_{{\cal N}=4}(\l)$ as a function of the 't Hooft coupling.

Remarkably, we find that the two--loop four--point function for the ABJM model can be obtained from the second order expansion of the same BDS--like ansatz where the four dimensional scaling function is substituted by the three dimensional one,
$f_{CS}(\lambda)$ as obtained from the conjectured asymptotic Bethe equations \cite{GV}.

Therefore, we make the conjecture that the all loop four--point amplitude is given by
\begin{equation}
\frac{\mathcal{A}_4}{\mathcal{A}_4^{tree}} = e^{Div + \frac{f_{CS}(\lambda)}{8}\left(\ln^2\left(\frac{s}{t}\right)+ \frac{4 \pi^2}{3} + 6 \ln^2 2\right)  + C(\lambda) }
\end{equation}
where now $\l$ is the ABJM coupling and $C(\l)$ is a scheme--dependent constant.

Since $f_{CS}(\l)$ is known up to order $\l^4$ \cite{LMMSSST,MOS}, we may predict the exact four--loop contribution to the four--point function (see eq. (\ref{conjecture})).\\

NOTE: When this work was already completed, a paper \cite{Chen:2011vv}
appeared, which has significant overlap. Although we draw the same
conclusions, we stress that our computation, being based on a direct
Feynman diagram approach is completely independent of that in
\cite{Chen:2011vv}, which makes use of generalized unitarity methods.

\section{ABJM model in ${\cal N}=2$ superspace}

An on--shell realization of  ${\cal N}=6$ supersymmetric ABJM models can be given in terms of
${\cal N}=2$ three dimensional superspace \cite{Klebanov}.  For $U(N) \times U(N)$
gauge group, the physical field content is organized into two vector multiplets $(V,\hat{V})$
in the adjoint representation of the first and the second $U(N)$'s,
coupled to chiral multiplets $A^i$ and $B_i$ carrying a fundamental index $i=1,2$ of a global $SU(2)_A \times SU(2)_B$ and in the bifundamental and antibifundamental representations of the gauge group, respectively.

The ${\cal N}=6$ supersymmetric action reads
\begin{equation}
 {\cal S} = {\cal S}_{\mathrm{CS}} + {\cal
    S}_{\mathrm{mat}}
  \label{eqn:action}
\end{equation}
with
\begin{eqnarray}
  \label{action}
  && {\cal S}_{\mathrm{CS}}
  =  \frac{K}{4\pi} \, \int d^3x\,d^4\theta \int_0^1 dt\: \Big\{  \Tr \Big[
  V \Db^\a \left( e^{-t V} D_\a e^{t V} \right) \Big]
  -   \Tr \Big[ \hat{V} \Db^\a \left( e^{-t \hat{V}} D_\a
    e^{t \hat{V}} \right) \Big]   \Big\}
  \non \\
  \non \\
  && {\cal S}_{\mathrm{mat}} = \int d^3x\,d^4\theta\: \Tr \left( \bar{A}_i
    e^V A^i e^{- \hat{V}} + \bar{B}^i e^{\hat V} B_i
    e^{-V} \right)
  \non \\
  &~& ~ + \frac{2\pi i}{K} \int d^3x\,d^2\theta\:
    \e_{ik} \e^{jl}  \, {\Tr (A^i B_j A^k B_l)} +  \frac{2\pi i}{K} \int d^3x\,d^2\bar{\theta}\:
  \e^{ik} \e_{jl} \, {\Tr (\bar{A}_i \bar{B}^j \bar{A}_k\bar{B}^l}  )
  \non\\
\end{eqnarray}
Here $K$  is an integer, as required by gauge invariance of the effective action.
In the perturbative regime we take $\l \equiv \frac{N}{K} \ll 1$.

The quantization of the theory can be easily carried on in superspace
after performing gauge fixing (for details, see for instance
\cite{BPS}). In momentum space and using Landau gauge, this
leads to gauge propagators
\begin{eqnarray}
&&  \langle V^a_{\, b}(1) \, V^c_{\, d}(2) \rangle
  =   \frac{4\pi}{K} \, \frac{1}{p^2} \,  \, \delta^a_d \, \d^c_b \times \Db^\a D_\a \, \delta^4(\th_1-\th_2) \nonumber \\
&&  \langle \hat V^{\bar{a}}_{\bar{b}} (1) \, \hat V^{\bar{c}}_{\bar{d}}(2) \rangle =
 - \frac{4\pi}{K} \,  \frac{1}{p^2} \, \,   \delta^{\bar{a}}_{\bar{d}} \, \d^{\bar{c}}_{\bar{b}} \times \Db^\a D_\a  \, \delta^4(\th_1-\th_2)
  \label{gaugeprop}
\end{eqnarray}
whereas the matter propagators are
\begin{eqnarray}
  &&\langle \bar A^{\bar{a}}_{\ a}(1) \, A^b_{\ \bar{b}}(2) \rangle
  = \frac{1}{p^2} \,\, \delta^{\bar{a}}_{\ \bar{b}} \, \delta^{\ b}_{a} \times  \delta^4(\th_1 - \th_2)
 \nonumber \\
  &&  \langle \bar B^a_{\ \bar{a}}(1) \, B^{\bar{b}}_{\ b}(2) \rangle =
   \frac{1}{p^2} \,\, \delta^a_{\ b} \, \delta^{\ \bar{b}}_{\bar{a}} \times \delta^4(\th_1 - \th_2)
\label{scalarprop}
\end{eqnarray}
where $a,b$ and $\bar{a}, \bar{b}$ are indices of the fundamental representation of the first and the second gauge groups, respectively.
The vertices employed in our two--loop calculation can be easily read from
the action (\ref{action}) and they are given by
\begin{eqnarray}
  \label{vertices}
 && \int d^3x\,d^4\theta\: \left[ \Tr ( \bar{A}_i V A^i) -  \Tr ( B_i V \bar{B}^i )
    + \Tr (  \bar{B}^i {\hat V} B_i ) -  \Tr ( A^i {\hat{V}}   \bar{A}_i ) \right]
    \non \\
  && \qquad + \frac{4\pi i}{K} \int d^3x\,d^2\theta\:
    \, \Big[ \Tr (A^1 B_1 A^2 B_2) -  \Tr (A^1 B_2 A^2 B_1)\Big] ~+~ {\rm h.c.}
\end{eqnarray}
We work in euclidean superspace with the effective action defined as $e^{\G} = \int e^S$.

Since the vector fields are not propagating, the only non--trivial amplitudes of the theory are the ones
involving matter external particles. In ${\cal N}=2$ superspace language this means having $A$, $B$ and their complex conjugates  as external superfields. Given the structure of the vertices, it is straightforward to
see that only amplitudes with an even number of external legs are non--vanishing. This is consistent
with the requirement for the amplitudes to be Lorentz and dilatation invariant \cite{HL}.

Each external scalar particle carries an on--shell momentum $p_{\a\b}$ ($p^2 =0$), an $SU(2)$ index
and color indices corresponding to the two gauge groups. We classify as
particles the ones carrying $(N, \bar{N})$ indices and antiparticles the ones carrying $(\bar{N}, N)$
indices. Therefore, $(A^i, \bar{B}^j)$ are {\em particles}, whereas  $(B_i, \bar{A}_j)$ are {\em antiparticles}.

We are interested in the simplest non--trivial amplitudes, that is four--point amplitudes. Without loosing
generality we consider the $(A B A B)$ superamplitude. All the other superamplitudes  can be obtained from this one by
$SU(4)$ R--symmetry transformations.

The color indices can be stripped out, as we can write
\beq
{\cal A}_4 \left( A^{a_1}_{\, \bar{a}_1}\,  B^{\bar{b}_2}_{\, b_2} \, A^{a_3}_{\, \bar{a}_3}
B^{\bar{b}_4}_{\, b_4} \right) =
\sum_{\s}
{\cal A}_4(\s(1),   \cdots , \s(4) ) \; \d^{a_{\s(1)}}_{b_{\s(2)}} \, \d^{\bar{b}_{\s(2)}}_{\bar{a}_{\s(3)}}
\, \d^{a_{\s(3)}}_{b_{\s(4)}} \d^{\bar{b}_{\s(4)}}_{\bar{a}_{\s(1)}}
\eeq
where the sum is over exchanges of even or odd sites between themselves.

\section{The four point amplitude at two loops}

We study four--point scattering amplitudes of the type $(A^i B_j A^k B_l)$, where the external $A,B$ particles carry outgoing momenta $p_1,\dots,p_4$ ($p_i^2=0$).  As usual, Mandelstam variables are defined by $s=(p_1+p_2)^2, t=(p_1+p_4)^2, u=(p_1+p_3)^2$.

\begin{figure}[h!]
    \centering
    \includegraphics[width=0.4\textwidth]{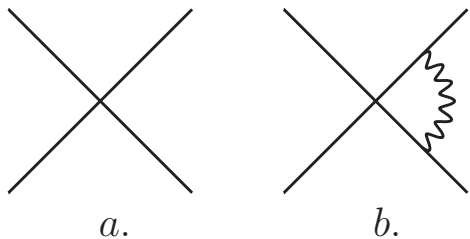}
    \caption{Diagrams contributing to the tree level and 1--loop four--point scattering amplitude. }
    \label{0set}
\end{figure}

At tree level the amplitude is simply given by the diagram in Fig. \ref{0set} (a) associated
to the classical superpotential in (\ref{vertices}). Its explicit expression is
\beq
{\cal A}_4^{tree}(A^i(p_1), B_j(p_2), A^k(p_3), B_l (p_4)) = \frac{2\pi i}{K}  \e^{ik} \e_{jl}
\eeq
At one loop it has been proved to vanish  \cite{ABM}. In ${\cal N}=2$  superspace language a symmetry argument shows that the only diagram that can be constructed (Fig. \ref{0set}$b$) leads to a vanishing contribution both off--shell and on--shell \cite{AndreaMati}.

\begin{figure}[h!]
    \centering
    \includegraphics[width=0.7\textwidth]{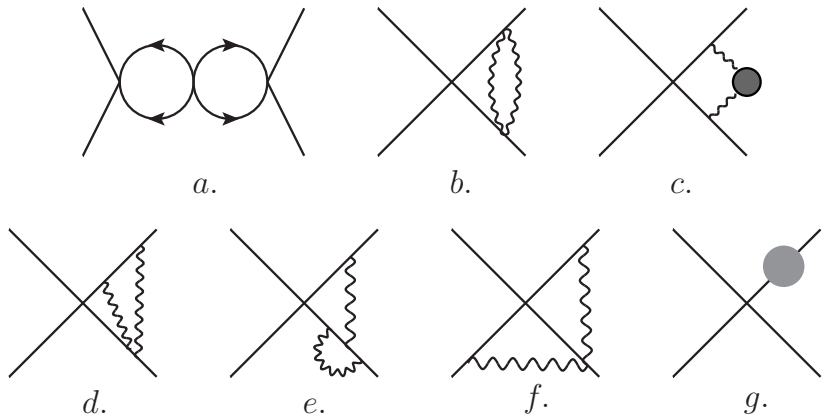}
    \caption{Diagrams contributing to the two--loop four--point scattering amplitude. The dark--gray blob represents one--loop corrections and the light--gray blob two--loop ones.}
    \label{1set}
\end{figure}

At two loops, in the planar sector, the amplitude can be read from the single trace part of the two--loop effective superpotential
\begin{eqnarray}
\label {effepotential}
&&\Gamma^{(2)}[A,B]=\int d^2\theta d^3p_1\dots d^3p_4\,(2\pi)^3\,\delta^{(3)}({\sum}_i p_i)\times\\
&&\frac{2\pi i}{K}\epsilon_{ik}\epsilon^{jl}\,\mathrm{tr}\left(A^i(p_1) B_j(p_2)A^k(p_3)B_l(p_4)\right)\sum\limits_{X=a}^{g}\mathcal{M}^{(X)}(p_1,\dots,p_4)
\end{eqnarray}
where the sum runs over the first six diagrams in Fig. \ref{1set}, plus the contribution from the 1P--reducible (1PR) graph in Fig. \ref{1set}$g$ where the bubble indicates the two--loop correction to the chiral propagator.

In (\ref{effepotential}) we have factorized the tree level expression, so that
$ \mathcal{M}^{(X)}(p_1,\dots,p_4)$ are contributions to ${\cal A}_4^{(2 \, loops)}/{\cal A}_4^{tree}$.

In order to evaluate the diagrams we fix the convention for the upper--left leg
to  carry momentum $p_1$ and name the other legs counterclockwise. The total contribution from every single graph is then given by summing over all possible permutations of the external legs accounting for the different scattering channels.

The momentum--dependent contributions in (\ref{effepotential}) are the product of a combinatorial factor times a sum of ordinary Feynman momentum integrals arising after performing D--algebra on each supergraph (details can be found in \cite{AndreaMati, BLMPS2}).

Massless scattering amplitudes  are affected by IR divergences. We deal with them by dimensional regularization, $d = 3 -2\e$, $\e<0$.

We begin by evaluating the simplest graph 2$a$. After performing D--algebra, its $s$--channel contribution
shown in Fig. \ref{1set} is given by a two--loop factorized Feynman integral  
\begin{equation}
\mathcal{D}^s_a=\mu^{4\epsilon}\int\frac{d^d k}{(2\pi)^d}\frac{d^d l}{(2\pi)^d}\frac{-(p_1+p_2)^2}{k^2\,(k+p_1+p_2)^2\, l^2\,(l-p_3-p_4)^2}=
-G[1,1]^2\left(\frac{\mu^2}{s}\right)^{2\epsilon}
\end{equation}
where $\mu$ is the mass scale of dimensional regularization and the $G$ function is defined by
\begin{equation}
G[a,b]=\frac{\Gamma(a+b-d/2)\Gamma(d/2-a)\Gamma(d/2-b)} {(4\pi)^{d/2}\Gamma(a)\Gamma(b)\Gamma(d-a-b)}
\end{equation} 
Taking into account all contributions of this type with color/flavor factors we obtain
\begin{equation}
\mathcal{M}^{(a)}=-(4\pi\lambda)^2 G[1,1]^2\left(\left(\frac{\mu^2}{s}\right)^{2\epsilon}+\left(\frac{\mu^2}{t}\right)^{2\epsilon}\right)=-3\zeta_2\lambda^2+\mathcal{O}(\epsilon)
\end{equation}

The contribution from diagram 2$b$, after D--algebra and with the particular assignment of momenta 
as in figure, is given by
\begin{equation} 
\mathcal{D}^{s_1}_b=\mu^{4\epsilon}\int\frac{d^d k}{(2\pi)^d}\frac{d^d l}{(2\pi)^d}\frac{2(p_3+p_4)^2}{l^2\,(l+k)^2\,(k-p_4)^2\,(k+p_3)^2}=
\frac{2 G[1,1] \Gamma(1+2\epsilon)\Gamma^2(-2\epsilon)}{(4\pi)^{d/2}\Gamma(1/2-3\epsilon)\,( s/\mu^2)^{2\epsilon}}
\end{equation}
Therefore, summing over all four contributions we get
\begin{equation}
\mathcal{M}^{(b)}=(4\pi\lambda)^2 \frac{ G[1,1] \Gamma(1+2\epsilon)\Gamma^2(-2\epsilon)}{(4\pi)^{d/2}\Gamma(1/2-3\epsilon)}\left(\left(\frac{\mu^2}{s}\right)^{2\epsilon}+\left(\frac{\mu^2}{t}\right)^{2\epsilon}\right)
\end{equation}
which is infrared divergent.

Diagram 2$c$, in contrast with the previous ones, is infrared divergent even when considered off--shell. This unphysical infrared divergence is cured by adding the 1PR diagram corresponding to two--loop self--energy corrections to the superpotential, depicted in Fig. \ref{1set}$g$. In fact, the contribution from this diagram, when the correction is on the $p_4$ leg, yields
\begin{equation}\label{gamba}
\mathcal{D}^{4}_g=-3\,G[1,1]\,G[1,3/2+\epsilon]\,(p_4^2)^{-2\epsilon}+2G[1,1]^2\,(p_4^2)^{-2\epsilon}
\end{equation}
The first term of this expression is infrared divergent even off--shell, but precisely cancels the infrared divergence of diagram 2$c$. The second term in (\ref{gamba}) comes from a double factorized bubble and is finite when $d\to 3$, but since we take the momenta to be on--shell before expanding in $\epsilon$, this piece vanishes on--shell. It turns out that after this cancelation between diagrams 2$c$ and 2$g$ the remainder is proportional to the integral corresponding to diagram 2$b$. Precisely, we have
\begin{equation}
\mathcal{M}^{(c)}+\mathcal{M}^{(g)}=-3\mathcal{M}^{(b)}
\end{equation}

Diagrams of type 2$d$ can be evaluated by using Mellin--Barnes techniques. Specifically, with the momenta assignment as in figure, the D--algebra gives
\begin{align}
\mathcal{D}^{s1}_d &=\mu^{4\epsilon}\int \frac{d^d k}{(2\pi)^d}\frac{d^d l}{(2\pi)^d}
\frac{Tr(\gamma_{\mu}\gamma_{\nu}\gamma_{\rho}\gamma_{\sigma})\,
p_4^{\mu}\,(p_3+p_4)^{\nu}\,(k+p_4)^{\rho}\,(l-p_4)^{\sigma}}{(k+p_4)^2\,(k-p_3)^2\,(k+l)^2\,(l-p_4)^2\,l^2}\\
&=-\frac{\Gamma^3(1/2-\epsilon)\Gamma(1+2\epsilon)\Gamma^2(-2\epsilon)}
{(4\pi)^{d}\Gamma^2(1-2\epsilon)\Gamma(1/2-3\epsilon)\,(s/\mu^2)^{2\epsilon}}
\end{align}
and summing over the eight permutations multiplied by the corresponding flavor/color factors we obtain 
\begin{equation}
\mathcal{M}^{(d)}=
-(4\pi\lambda)^2\frac{2\Gamma^3(1/2-\epsilon)\Gamma(1+2\epsilon)\Gamma^2(-2\epsilon)}
{(4\pi)^{d}\Gamma^2(1-2\epsilon)\Gamma(1/2-3\epsilon)}\left(\left(\frac{\mu^2}{s}\right)^{2\epsilon}+\left(\frac{\mu^2}{t}\right)^{2\epsilon}\right)
\end{equation}

Using the identities derived in \cite{AndreaMati} it is possible to write diagram 2$e$ as a combination of diagrams 2$b$ and 2$d$ plus a double factorized bubble which can be dropped when working on--shell. We find
\begin{equation}
\mathcal{M}^{(e)}=2\mathcal{M}^{(d)}+4\mathcal{M}^{(b)}
\end{equation}

The most complicated contribution comes from diagram 2$f$, which involves a nontrivial function of the ratio $s/t$ of kinematic invariants. Surprisingly, after some cancelations it turns out to be finite. The D--algebra for the specific choice of the external momenta as in figure results in the Feynman integral
\begin{equation}
\mathcal{D}^{234}_{f}=\mu^{4\epsilon}\int \frac{d^d k}{(2\pi)^d}\frac{d^d l}{(2\pi)^d}
\frac{-Tr(\gamma_{\mu}\gamma_{\nu}\gamma_{\rho}\gamma_{\sigma})\,
p_4^{\mu}\,p_2^{\nu}\,k^{\rho}\,l^{\sigma}}{k^2\,(k-p_2)^2\,(k+l+p_3)^2\,(l-p_4)^2\,l^2}
\end{equation}
which after taking the on--shell limit can be expressed exactly as a single one--fold Mellin--Barnes integral which is finite in the limit $\epsilon\to 0$
\begin{align}
&\mathcal{D}^{234}_{f}=\frac{(1+s/t)\Gamma^3(1/2-\epsilon)}{(4\pi)^d\Gamma^2(1-2\epsilon)\Gamma(1/2-3\epsilon)(t/\mu^2)^{2\epsilon}}\times\\
\times\int\limits^{+i\infty}_{-i\infty} \frac{d{\bf v}}{2\pi i}\Gamma(-{\bf v}) & \Gamma(-2\epsilon -{\bf v})
\Gamma^{*}(-1-2\epsilon-{\bf v})\Gamma^2(1+{\bf v})\Gamma(2+2\epsilon+{\bf v})\left(\frac{s}{t}\right)^{{\bf v}}
\end{align}
Taking into account the four permutations, flavor/color factors and expanding in $\epsilon$ we get
\begin{equation}
\mathcal{M}^{(f)}=\lambda^2\left(\tfrac{1}{2}\ln^2(s/t)+3\zeta_2\right)+\mathcal{O}(\epsilon)
\end{equation}

Collecting all the partial results, after some algebra we may reduce the result to the following compact form
\begin{equation}
\label{result}
\mathcal{M}^{(2)} \equiv \frac{\mathcal{A}_4^{(2 \, loops)}}{\mathcal{A}^{tree}_4} = \lambda^2\, \left[-\frac{( s/\mu'^2)^{-2\epsilon}}{(2\, \epsilon)^2}-\frac{(t/\mu'^2)^{-2\epsilon}}{(2\, \epsilon)^2}+\frac12\,\ln^2 \left(\frac{s}{t}\right)+{\cal C}+
\mathcal{O}(\epsilon)\right]
\end{equation}
where $\mu'^2=8\pi e^{-\gamma}\,\mu^2$, and ${\cal C}$ is a constant given by ${\cal C}=4\zeta_2+3\ln^2 2$.\footnote{We note that the analytical value of the constant term matches the numerical result of \cite{Chen:2011vv}.}

If we rotate to Minkowski spacetime with mostly minus signature and write the Mandelstam variables in terms of the dual ones, $s = -x_{13}^2, t=-x_{24}^2$,
up to a (scheme--dependent) constant, our result matches the expression (\ref{WL}) for the two--loop expansion of a light--like Wilson loop, once we  have identified the UV and IR rescaled regulators of the Wilson loop and scattering amplitude, as $1/\mu_{WL}^2 = \mu'^2$.

\section{Dual conformal invariance}

The two--loop result (\ref{result}) for the four--point amplitude in ABJM theories has the same functional
structure as the one--loop correction to the four--point amplitude in four dimensional ${\cal N}=4$ SYM theory \cite{BDS, Drummond:2007aua}, provided that we rescale $\e \to 2\e$ there.

In the ${\cal N}=4$ SYM case, the perturbative results for planar MHV scattering amplitudes can be expressed as linear combinations of scalar integrals that are off--shell finite in four dimensions and dual conformal invariant \cite{Drummond:2007aua}.  Precisely, once written in terms of dual variables, $p_i = x_{i+1} - x_i$, the integrands times the measure are invariant under translations, rotations, dilatations and special conformal transformations. In particular, invariance under inversion, $x^{\mu} \to x^{\mu}/x^2$, rules out bubbles and triangles and up to two loops, only square--type diagrams appear.

Dual conformal invariance is broken on--shell by IR divergences that require introducing a mass regulator. Therefore, conformal Ward identities acquire an anomalous contribution \cite{Drummond:2007au}.

A natural question  which arises is whether the two--loop result (\ref{result}) for three dimensional ABJM models exhibits dual conformal invariance.

In order to answer this question, we concentrate on the momentum integrals associated to the four diagrams in Fig. \ref{1set} which are the ones that eventually combine to lead to the final result (\ref{result}).
We study their behavior under dual conformal transformations when evaluated off--shell and in three dimensions.

We first rewrite their expressions in terms of dual variables and then perform
conformal transformations, the only non--trivial one being the inversion.

Since under inversion $x_{ij}^2 \to \frac{x_{ij}^2}{x_i^2 x_j^2}$ and $d^dx_i \to \frac{d^dx_i}{(x_i^2)^d}$,
it is easy to realize that, while in four dimensions the elementary invariant building block integrands are squares, in three dimensions they should be triangles.
Therefore, it is immediate to conclude that the integrands associated to diagrams $\ref{1set}a-\ref{1set}b$ cannot be invariant, since they contain bubbles. On the other hand, diagrams  $\ref{1set}d-\ref{1set}f$  contain triangles but also non--trivial numerators which concur to make the integrand non--invariant under inversion.

Despite dual conformal invariance seems not to be a symmetry of the integrals arising from our Feynman diagram approach, in the previous Section we have showed that the on--shell amplitude, when written in dual space, has the same functional form of the light--like Wilson loop.
As a consequence, on--shell the amplitude should possess dual conformal invariance, since Wilson loops inherit the ordinary conformal invariance of the ABJM theory, even though anomalously broken by UV divergences.

As a consequence, it should be possible to rewrite expression (\ref{result}) for the on--shell amplitude as a linear combination of scalar integrals which are off--shell finite in three dimensions and manifestly dual conformal invariant at the level of the integrands\footnote{Note added: This task has been actually accomplished in \cite{Chen:2011vv}, where an explicit basis of dual conformal integrals has been determined on which the amplitude can be expanded.}.

\section{A conjecture for the all--loop four--point amplitude}

Our result in (\ref{result}) provides the first non--trivial contribution to the four--point scattering amplitude in the ABJM theory. The analogue quantity in four dimensional $\mathcal{N}=4$ SYM  has been extensively studied and an all--loop iteration conjecture for it has been given in \cite{BDS, Bern:2008ap}. The result may be schematically written as \\
\begin{equation}\label{BDScon}
\frac{\mathcal{A}}{\mathcal{A}^{tree}} = e^{Div + \frac{f_{\mathcal{N}=4}(\lambda)}{8}\left(\ln^2\left(\frac{s}{t}\right)+ \frac{4 \pi^2}{3}\right)  + C(\lambda) }
\end{equation}
where $f_{\mathcal{N}=4}(\lambda)$ is the scaling function of $\mathcal{N}=4$ SYM in terms of the  't Hooft coupling $\lambda=g^2 N$, the constant $C(\lambda)$ is independent of the kinematic variables and the IR divergent contributions are grouped in the first term.

It would be interesting to check whether a similar resummed expression may hold for scattering amplitudes in the three--dimensional case.
Although we only computed the first non--trivial perturbative order for the amplitude,  still we have some indications that this could be the case.

At first, comparing the conjectured form of the asymptotic all loop Bethe equations for $\mathcal{N}=4$ SYM and ABJM theory, Gromov and Vieira noticed \cite{GV} that the scaling functions of the two theories should be related as
\begin{equation}\label{3d4d}
f_{CS}(\lambda)= \left. \frac{1}{2}f_{\mathcal{N}=4} (\lambda)\right|_{\frac{\sqrt{\lambda}}{4 \pi}\rightarrow h(\lambda)}
\end{equation}
where $h(\lambda)$ is the interpolating function of the magnon energy dispersion relation. The first perturbative orders of  $h(\lambda)$ have been computed at both weak \cite{LMMSSST,MOS} and strong coupling \cite{Gaiotto:2008cg}--\cite{Astolfi:2011ju}. The weak coupling expansion
\begin{equation} \label{hfun}
 h^2(\lambda)  =  \lambda^2 - 4 \zeta_2 \,\lambda^4 + \mathcal{O}(\lambda^6)  \hspace{1.8cm}   \lambda \ll 1
\end{equation}
can be combined, using (\ref{3d4d}),  with the known expansion of the 4d scaling function $f_{\mathcal{N}=4}(\lambda)= \lambda/2\pi^2 -1/96\pi^2 \lambda^2 + O(\lambda^3)$ . We are then able to write explicitly the 3d scaling function up to order $\lambda^4$
\begin{equation} \label{3df}
f_{CS}(\lambda) = 4 \lambda^2 - 4 \pi^2 \lambda^4 + O(\lambda^6)
\end{equation}
Assuming (\ref{BDScon}) to hold also in the three dimensional case with the very same constant coefficients and plugging (\ref{3df}) in it,  after expanding at order $\lambda^2$, we curiously find an exact correspondence with the result we explicitly computed in (\ref{result}). This suggests that for the three dimensional case, provided we use the correct scaling function, a completely analogous resummation may take place to give an expression for the amplitude of the form
\begin{equation}\label{con}
\frac{\mathcal{A}_4}{\mathcal{A}^{tree}_4} = e^{Div + \frac{f_{CS}(\l)}{8}\left(\ln^2\left(\frac{s}{t}\right)+ 8\zeta_2 + 6\,\ln^2 2\right)  + C(\lambda) }
\end{equation}

If this is the case,  using (\ref{3df}), we may predict the next non--trivial order for the finite remainder $F_4^{(4)}$ (in the notation of  \cite{BDS}) of the four--point scattering amplitude 
\beq
\label{conjecture}
F_4^{(4)} = \frac{\l^4}{8} \ln^4
\left(\frac{s}{t}\right) + \l^4 \left(\frac{3}{2}\ln^2 2- \zeta_2\right) \ln^2\left(\frac{s}{t}\right) + {\rm Consts}
\eeq
A direct check of this prediction, either with a 4--loop scattering amplitude computation or using the duality with Wilson loops, could confirm the conjectured exact expression in (\ref{con}).

\section{Conclusions}

We briefly summarize the main results of this paper and discuss future developments.

For three dimensional ABJM superconformal models, in a ${\cal N}=2$ superspace setup, we have computed the planar, two--loop corrections to the chiral $(ABAB)$ four--point superamplitude. We performed the calculation by a direct Feynman diagram approach, in a manifestly supersymmetric formalism. We have found a non--vanishing result which perfectly agrees with the two--loop result for a light--like four--polygon Wilson loop.
This result represents the first non--trivial evidence of an amplitude/WL duality working in three dimensional superconformal theories and confirms the conjectured duality which seemed to arise trivially at one loop.

Its functional structure resembles the one--loop planar four--point amplitude for ${\cal N}=4$ SYM theory in four dimensions. As in that case, it can be obtained from a BDS--like ansatz for the all--loop amplitude where the scaling function of four dimensions is substituted by the three dimensional one, as predicted by the conjectured Bethe equations.

For ${\cal N}=4$ SYM theory the structure of the four--point BDS ansatz has been verified also at strong coupling \cite{AM}. It would be interesting to check whether applying the recipe of \cite{AM} for computing scattering amplitudes at strong coupling to the ABJM case, the result agrees with a three dimensional version of the BDS ansatz. From our weak coupling computation we expect this to be the case, at least at four points. A preliminary discussion will be given in \cite{BLMPS2}.

An important question to be addressed is whether and how dual conformal invariance plays a role in three dimensional models. By explicit calculations, which do not assume dual conformal invariance a priori, we showed that the four--point on--shell amplitude is dual to the four--cusps light--like Wilson loop. This hints at the invariance of the result under dual conformal invariance, even though this symmetry is not manifest in our Feynman diagram approach. We will report on this issue more extensively in \cite{BLMPS2}.

\end{document}